\documentclass[12pt]{aastex}
\usepackage{emulateapj5}
\begin{document}
\shortauthors{McKinney \& Gammie}
\shorttitle{Numerical Viscous Accretion Flows}

\title{Numerical Models of Viscous Accretion Flows Near Black
Holes}

\author{Jonathan C. McKinney$^{1}$ and Charles F. Gammie$^{1,2}$}

\altaffiltext{1}{Physics Department and Center for Theoretical 
Astrophysics, University of Illinois at Urbana-Champaign, 
Loomis Laboratory, 1110 W.Green St. Urbana,IL 61801}

\altaffiltext{2}{Department of Astronomy and National Center for 
Supercomputing Applications (NCSA)}

\email{jcmcknny@uiuc.edu, gammie@uiuc.edu}

\begin{abstract}

We report on a numerical study of viscous fluid accretion onto a
black hole.  The flow is axisymmetric and uses a pseudo-Newtonian
potential to model relativistic effects near the event horizon.
The numerical method is a variant of the ZEUS code.  As a test of
our numerical scheme, we are able to reproduce results from
earlier, similar work by Igumenshchev and Abramowicz and Stone et
al.  We consider models in which mass is injected onto the grid as
well as models in which an initial equilibrium torus is accreted.
In each model we measure three ``eigenvalues'' of the flow: the
accretion rate of mass, angular momentum, and energy.  We find that
the eigenvalues are sensitive to $r_{in}$, the location of the
inner radial boundary.  Only when the flow is always supersonic on
the inner boundary are the eigenvalues insensitive to small changes
in $r_{in}$.  We also report on the sensitivity of the results to
other numerical parameters.

\end{abstract}


\maketitle

\section{Introduction}

Black hole accretion flows are the most likely central engine for
quasars and active galactic nuclei (AGN) \citep{z64,salpeter64}.
As such they are the subject of intense astrophysical interest and
speculation.  Recent observations from XMM-Newton, Chandra, Hubble,
VLBA, and other ground- and space-based observatories have expanded
our understanding of the time variability, spectra, and spatial
structure of AGN.  Radio interferometry, in particular, has been
able to probe within a few hundred gravitational radii ($G M/c^2$)
of the central black hole, e.g.  \citet{lo98,junor99,doel2001}.
Despite these observational advances, only instruments now in the
concept phase will have sufficient angular resolution to spatially
resolve the inner accretion disk \citep{rees2001}.  And so there
remain fundamental questions that we can only answer by folding
observations through models of AGN structure.

All black hole accretion flow models require that angular momentum
be removed from the flow in some way so that material can flow
inwards.  In one group of models, angular momentum is removed
directly from the inflow by, e.g., a magneto-centrifugal wind
\citep{bp82}.  Here we will focus on the other group of models in
which angular momentum is diffused outward through the accretion
flow.

It has long been suspected that the diffusion of angular momentum
through an accretion flow is driven by turbulence.  The $\alpha$
model \citep{ss73} introduced a phenomenological shear stress into
the equations of motion to model the effects of this turbulence.
This shear stress is proportional to $\alpha P$, where $\alpha$ is
a dimensionless constant and $P$ is the (gas or gas $+$ radiation)
pressure.  This shear stress permits an exchange of angular
momentum between neighboring, differentially rotating layers in an
accretion disk.  In this sense it is analogous to a viscosity (see
also \cite{lbp74}) and is often referred to as the ``anomalous
viscosity.''

The $\alpha$ model artfully avoids the question of the origin and
nature of turbulence in accretion disks.  This allows useful
estimates to be made absent the solution to a difficult, perhaps
intractable, problem.  Recently, however, significant progress has
been made in understanding the origin of turbulence in accretion
flows.  It is now known that, in the magnetohydrodynamic (MHD)
approximation, an accreting, differentially rotating plasma is
destabilized by a weak magnetic field \citep{bh91,hb91}.  This
magneto-rotational instability (MRI) generates angular momentum
transport under a broad range of conditions.  Numerical work has
shown that in a plasma that is fully ionized, which is likely the
case for the inner regions of most black hole accretion flows, the
MRI is capable of sustaining turbulence in the nonlinear regime
\citep{hb91,h95,h00,hk01}.

Studies of unmagnetized disks have greatly reduced the probability
that a linear or nonlinear hydrodynamic instability drives disk
turbulence.  While there are known global hydrodynamic
instabilities that could in principle initiate turbulence, these
have turned out to saturate at low levels or require conditions
that are not relevant to an accretion disk near a black hole.  As
of this writing, no local, linear or nonlinear hydrodynamic
instabilities that transport angular momentum outwards are known to
exist in Keplerian disks \citep{bh98}.

Work on magnetized disks has now turned to global numerical models.
These are possible thanks to advances in computer hardware and
algorithms.  Recent work by \citet{h00,h01}, \citet{sp01}, and
\citet{hk01} considers the evolution of inviscid, nonrelativistic
MHD accretion flows in two or three dimensions.  Some of this work
uses a pseudo-Newtonian, or \citet{pw80}, potential as a model for
the effects of strong-field gravity near the event horizon.  

Other work on global models has considered the equations of
viscous, compressible fluid dynamics as a model for the accreting
plasma \citep{ia99,spb99,ia00,ian00}.  The viscosity is meant to
mock up the effect of small scale turbulence, presumably generated
by magnetic fields, on the large scale flow.  In light of work on
numerical MHD models, this may seem like a step backwards.  The MHD
models, however, are computationally expensive and introduce new
problems with respect to initial and boundary conditions.  It
therefore seems reasonable to investigate the less expensive
$\alpha$ based viscosity models.  In this paper we investigate
axisymmetric, numerical, viscous inflow models.

This work was motivated by the earlier work of \citet{ia99,ia00} and
\citet{spb99}, hereafter referred to as IA99, IA00, and SPB99,
respectively (IA99 and IA00 are collectively referred to as IA, in
which case SPB99 is simply referred to as SPB).  These authors studied
similar viscous inflow models yet found different radial scaling laws
for radial velocity, density, and angular momentum.  They also found
different values for the accretion rate of mass and angular momentum.
They used different experimental designs, however.  We set out to
discover whether the results from these authors differed due to
numerical methods or model parameters.

Along the way, we took a systematic approach to studying numerical
parameters and boundary conditions.  One particular point of
concern, which will be described in greater detail below, is the
inner boundary condition.  This lies in the energetically dominant
portion of the flow, so errors there can potentially corrupt the
entire model.   In this paper we show that aspects of results
presented by other researchers are sensitive to model and numerical
parameters.  These results should be useful to others contemplating
large-scale numerical models of black hole accretion.

The paper is organized as follows.  In \S 2 we discuss our models.
In \S 3 we discuss numerical methods.  In \S 4 we discuss a fiducial
solution and results from a survey of other solutions.  In \S 5 we
summarize our results.

\section{Model}

We are interested in modeling the plasma within a few hundred $G
M/c^2$ of a black hole.  We will consider only axisymmetric
models (the work of \cite{ian00} suggests that 2D and 3D viscous
models give similar results).  Throughout we use standard
spherical polar coordinates $r$, $\theta$, and $\phi$.

We solve numerically the axisymmetric, nonrelativistic equations
of compressible hydrodynamics in the presence of an anomalous
stress $\mathbf{\Pi}$, which is meant to model the effects of
small-scale turbulence on the mean flow.  The governing
equations then express the conservation of mass
\begin{equation}
\frac{D\rho}{Dt} + \rho(\nabla\cdot\mathbf{v}) = 0,
\end{equation}
momentum,
\begin{equation}\label{EOM}
\rho\frac{D\mathbf{v}}{Dt} = -\nabla{P} - \rho\nabla\Psi 
	- \nabla\cdot\mathbf{\Pi},
\end{equation}
and energy,
\begin{equation}
\frac{Du}{Dt} = -(P+u)(\nabla\cdot\mathbf{v})+\Phi.
\end{equation}
Here, as usual,
$D/Dt\equiv\partial/\partial{t}+\mathbf{v}\cdot\nabla$ is the
Lagrangian time derivative, $\rho$ is the mass density,
$\mathbf{v}$ is the velocity, $u$ is the internal energy density,
$P$ is the pressure, and $\Psi$ is the gravitational potential.  
The dissipation function $\Phi$ is given by the product of the
anomalous stress tensor $\mathbf{\Pi}$ with the rate-of-strain
tensor $\mathbf{e}$ (given explicitly in the Appendix)
\begin{equation}
\Phi=\Pi_{ij} e^{ij},
\end{equation}
(sum over indices) where the anomalous stress tensor is the
term-by-term product
\begin{equation}
\Pi_{ij} = -2\rho\nu e_{ij} S_{ij},
\end{equation}
(no sum over indices) where $S_{ij}$ is a symmetric matrix filled
with $0$ or $1$ that serves as a switch for each component of the
anomalous stress.  The equation of state is
\begin{equation}
P = (\gamma - 1) u.
\end{equation}

For the gravitational potential we use the pseudo-Newtonian potential
of \citet{pw80}: $\Psi = -GM/(r - r_g)$ (here $r_g \equiv 2 G M/c^2$).
This potential reproduces features of the orbital structure of a
Schwarzschild spacetime, including an innermost stable circular orbit
(ISCO) located at $r = 6 G M/c^2$.  In a few cases we use the
Newtonian potential $\Psi = -GM/r$ for comparison with others' work
(IA and SPB exclusively use a Newtonian potential).

We must now make some choices for the anomalous stress tensor.  One
might argue on very general grounds for a Navier-Stokes prescription,
and indeed IA and we use a prescription where all elements of
$\mathbf{S}$ are $1$ (the ``IA prescription'').  SPB, on the other
hand, use the Navier-Stokes prescription with all components zero
except $S_{r\phi}$, $S_{\theta\phi}$, $S_{\phi r}$, and
$S_{\phi\theta}$ (the ``SPB prescription'').  SPB justify this choice
by arguing that it more appropriately models MHD turbulence;
this was later supported by results presented in \citet{sp01}.

We must also choose a viscosity coefficient.  We consider three
different prescriptions: one similar to those chosen by IA; a second
viscosity coefficient similar to that chosen by SPB; and a third,
similar form that vanishes rapidly near the poles.  Explicitly,
\begin{equation}
\nu = \alpha (c_s^2/\Omega_K),
\end{equation}
\begin{equation}
\nu = \alpha (\rho/\rho_0) \Omega_0 r_0^2,
\end{equation}
\begin{equation}
\nu = \alpha (c_s^2/\Omega_K) {\sin}^{3/2}(\theta),
\end{equation}
are the IA, SPB, and MG prescriptions, respectively, where
$c_s=\sqrt{\gamma P/\rho}$ is the sound speed, and
\begin{equation}
\Omega_K^2 \equiv {1\over{r}}{\partial{\Psi}\over{\partial{r}}} =
{GM\over{r (r - r_g)^2}}
\end{equation}
is the ``Keplerian'' angular velocity.  Here $\rho_0$ and $\Omega_0$
are values of $\rho$ and $\Omega$ at a fiducial radius $r_{0}$.  

The choice of viscosity coefficient for IA and MG is based, as usual, on
dimensional arguments.  SPB99's choice focuses the viscosity where most
of the matter is, a numerical convenience.  Our MG prescription is a
small modification of the IA prescription to concentrate the viscosity
toward the equator.  These choices are to a large extent arbitrary,
although one might attempt to motivate the choice by comparison with MHD
simulations, as do \citet{sp01}.  Nevertheless, some dynamical
properties of MHD turbulence, such as the elastic properties that
produce magnetic tension and hence Alfven waves, can never be modeled
with a viscosity.  

The model has boundaries at $\theta = \pm \pi/2$ and at $r = r_{in},
r_{out}$.  At the $\theta$ boundaries we use the usual polar axis
boundary conditions.  At the radial boundaries we use ``outflow''
boundary conditions; ideally these boundary conditions should be
completely transparent to outgoing waves.

Fuel for the accretion flow must be provided either in the initial
conditions or continuously over the model evolution.  Some global
numerical accretion flow models have started with an equilibrium
torus.  Examples include \citet{h91}, \citet{h95}, \citet{h00},
\citet{h01}, \citet{hbs01}, and \citet{hk01}.  Others have started
with an initial configuration of matter that is not in equilibrium.
For example, matter may be placed in orbit about the black hole,
but with sub-Keplerian angular momentum, so that once the
simulation commences it immediately falls toward the hole.
Examples of this approach include \citet{h84}, \citet{koide99},
\citet{koide00}, and \citet{mku2001}.  This approach may enhance
transients associated with the choice of initial conditions, although
it can also be physically well motivated, as in studies of
core-collapse supernovae.  One can also inject fluid continuously onto
the computational grid over the course of the evolution.  Examples of
this include IA99 and IA00.  One might also use an inflow boundary
condition at the outer radial edge of the grid, as in \citet{bbr2001}.
The main advantage of injection models is that they allow one to
achieve a steady, or statistically steady, state.  In this paper we
will consider only the equilibrium tori and on-grid injection models.

The equilibrium tori \citep{pp84} are steady-state solutions to the
equations of inviscid hydrodynamics.  They assume a polytropic
distribution of mass and internal energy, $P = K \rho^{\gamma}$,
and a power-law rotation profile, $\Omega\propto
(r\sin{\theta})^{-q}$.  The radial and meridional components of the
velocity vanish.  There are 5 parameters that describe the torus:
(1) the location of the torus pressure maximum $r_{0}$ ; (2) the
location of the innermost edge of the torus, $r_{t,in}<r_{0}$; (3)
the maximum value of the density, $\rho_{0} = \rho(r_0)$; (4) the
angular velocity gradient $q =d\ln{\Omega}/d\ln{R}$; and (5) the
entropy constant $K$.

On-grid injection adds matter to the model at a constant rate.  The
matter is injected with a non-zero specific angular momentum and with
zero radial or meridional momentum in a steady pattern
$\dot{\rho}(r,\theta)$ which is typically symmetric about the equator.
Parameters for this scheme include: (1) a characteristic radius for
injection $r_{inj}$; (2) the rate of mass injection $\dot{M}_{inj}$;
(3) the specific angular momentum of the injected fluid, $v_{\phi}=
f_1 r \Omega_K$.  We usually set $f_1 = 0.95$, so that the fluid
circularizes near $r_{inj}$.  This restricts transients associated
with circularization to the outer portions of the computational
domain; (4) the internal energy of the injected fluid, $u= f_2
\rho\Psi$.  We always set $f_2 = 0.2$ so that the fluid is marginally
bound, i.e. has Bernoulli parameter $Be \equiv (1/2) v^2 +
c_s^2/(\gamma-1) + \Psi < 0$.

One must also choose the injection pattern $\dot{\rho}(r,\theta)$.
IA99 choose a radially narrow region, but do not explicitly give
$\dot{\rho}(r,\theta)$.  We use a Gaussian, but found that none of the
results are sensitive to the precise profile.  The accretion rate of
mass, energy, and angular momentum are insensitive to large changes in
the size of the injection region except for the extreme cases of
filling the entire $\theta$ width or injecting in 2 locations.  These
extreme cases are sufficiently different to be referred to as a
completely different model; they lead to a qualitative change in the
flow.  For example, a full range $\theta$ injection region has matter
that will collide with any outflow at the poles.  A bipolar injection
leads to an equatorial outflow.  Our models have radial width
$\sigma_r = 0.05 (r_{in} + r_{out})/2$ and $\sigma_\theta = \pi/8$.

\section{Numerical Methods}

Our numerical method is based on ZEUS-2D \citep{sn92} with the
addition of an explicit scheme for the viscosity.  ZEUS is an
operator-split, finite-difference algorithm on a staggered mesh that
uses an artificial viscosity to capture shocks (in addition to the
anomalous viscosity in equations [\ref{EOM}]).  This algorithm
guarantees that momentum and mass are conserved to machine precision.
Total energy is conserved only to truncation error, so total energy
conservation is useful in assessing the accuracy of the evolution.

The inner and outer radial boundary conditions are implemented by
copying primitive variable values ($\rho$, $u$, and $\mathbf{v}$)
from the last zone on the grid into a set of ``ghost zones''
immediately outside the grid.  Inflow from outside the grid is
forbidden; we set $v_r(r_{in})=0$ if $v_r(r_{in})>0$ and
$v_r(r_{out})=0$ if $v_r(r_{out})<0$.  Since we expect inflow on
the inner boundary, this switch should seldom be activated.  We
have found that frequent activation of the switch is usually an
indication of a numerical problem.

We use a radial grid uniform in $\log(r - r_g)$.  We require that
$dr(r)/(r_{in}-r_g) \le 1/4$ so that the structure of the
pseudo-Newtonian potential is well resolved.  The grid is uniform
in $\theta$.  The grid has $N_r \times N_\theta$ zones.

\subsection{Numerical Treatment of Low Density Regions}

Like many schemes for numerical hydrodynamics, ZEUS can tolerate
only a limited dynamic range in density.  It is therefore
necessary to impose a density minimum $\rho_{fl}$ to avoid small
or negative densities.  Our procedure for imposing the floor is
equivalent to adding a small amount of mass to the grid every
time the floor is invoked.  Mass is added in such a way that
momentum is conserved.  To monitor the effect of the density
floor, we track the rate of change of total mass and total
energy (from kinetic energy change) due to this procedure,
$\dot{M}_{fl}$ and $\dot{E}_{fl}$.

We set $\rho_{fl}=10^{-10}\dot{M}_{inj} c^3 /(GM)^2$ for injection
runs and $\rho_{fl}=10^{-5}\rho_{0}$ for torus runs.  Lower values
for $\rho_{fl}$ do not lead to a significant change in the
solution.  Larger values of $\rho_{fl}$ give $\dot{M}_{fl} \sim
\dot{M}$, the accretion rate through the inner boundary.  The
atmosphere also becomes more massive and begins to affect torus
stability-- vertical oscillations are excited in the inner disk by
a Kelvin-Helmholtz like instability.  This should be avoided.

We must also surround the torus in a low density atmosphere in the
initial conditions.  The density of the atmosphere is $\rho_{fl}$ and
the internal energy density is $u = U_{o}\rho\Psi$, where $U_{o}$ is a
constant fraction of order unity (e.g. IA and we choose $U_{o}=0.2$).
The addition of the atmosphere has no effect on the solution since the
mass source's evolution eventually dominates the flow everywhere.
SPB99 choose a different method of constructing the initial atmosphere
but obtain late-time results that are similar to ours.

It is also necessary to impose a floor $u_{fl}$ on the internal energy
density.  This we take to be the minimum value of $u$ in the initial
atmosphere.  As for the mass, we track the rate of change of total
energy due to the internal energy floor, that along with the kinetic
energy is included in $\dot{E}_{fl}$.

\subsection{Diagnostics}

Global numerical simulations of accretion flows are complicated; it
is possible to measure many quantities associated with the flow.
Some are astrophysically relevant, and some are not.  In our view
particular interest attaches to the time-averaged flux of mass,
energy, and angular momentum through the inner boundary.
Physically, these are directly related to the luminosity of the
accretion flow and the rate of change of mass and angular momentum
of the central black hole.  As described by \cite{np93}, these are
in a sense the nonlinear ``eigenvalues'' of the model.

The mass accretion rate is
\begin{equation}
\dot{M} =  \int\limits_{S}\rho\mathbf{v}\cdot d\mathbf{S},
\end{equation}
where $S$ is the inner radial surface of the computational domain.
The total energy accretion rate is
\begin{equation}
\dot{E} = \int\limits_{S}((\frac{1}{2}v^2+h+\Psi)\rho\mathbf{v}+\mathbf{\Pi}\cdot\mathbf{v})\cdot d\mathbf{S},
\end{equation}
where $h=(u+P)/\rho=\gamma u/\rho$ is the specific enthalpy with our
equation of state.  The angular momentum accretion rate is
\begin{equation}\label{angmomflux}
\dot{L} = \int\limits_{S} r\sin{\theta}(\rho \mathbf{v} v_{\phi}+\mathbf{\Pi}\cdot \hat{\phi}) \cdot  d\mathbf{S}.
\end{equation}
It is also sometimes useful to focus on the reduced eigenvalues $l =
\dot{L}/\dot{M}$ and $e = \dot{E}/\dot{M}$.  These value of mass,
energy, and angular momentum are recorded at about 2 grid zones away
from $R_{in}$.  This avoids any error that may occur when evaluating
directly on the boundary where the inflow boundary condition is
applied.

We also track volume-integrated quantities, the flux of mass,
energy, and angular momentum across all boundaries, and floor added
quantities in order to evaluate the consistency of the results.
Mass and angular momentum are conserved to machine precision,
although ``machine precision'' implies a surprisingly large random
walk in the integrated quantities over the full integration because
the calculation requires millions of timesteps.

Total energy is conserved to truncation error, not machine
precision, and thus is a useful check on the quality of the
simulation.  Total energy conservation implies
\begin{equation}
\dot{E}_{err} = \dot{E}_{vol}+\dot{E}+\dot{E}_{out}-\dot{E}_{fl},
\end{equation}
where $\dot{E}_{vol}$ is the rate of change of the volume integrated
total energy, $\dot{E}_{out}$ is the flux of total energy through the
outer radial boundary, and $\dot{E}_{fl}$ is the rate of total energy
added due to the kinetic energy change (because of the mass density
floor) and internal energy density floor.  Ideally, $\dot{E}_{err}=0$.
Truncation errors can (and do) lead to cumulative, rather than random,
changes in the total energy.  A useful gauge of the magnitude of these
errors is $\dot{E}_{err}/E$.  For all runs we performed the error rate
is within $10\%$ of $10^{-5} c^3/GM$ for a torus run and within $10\%$
of $10^{-4} c^3/GM$ for a viscous injection run.

\subsection{Code Tests}

Our version of ZEUS reproduces all hydrodynamic test results from
\citet{sn92}, including their spherical advection and Sod shock
tests.  We also find excellent agreement with steady spherical
accretion solutions, i.e. Bondi flow /citep{bondi52}.  An inviscid
equilibrium torus run also persists for many dynamical times with
insignificant deviations from the initial conditions.

We have parallelized our code using the MPI message passing
library.  On the Origin 2000 at NCSA we are able to achieve about
$2.5\times 10^7$ zone updates per second using 240 CPUs, or about
$35$ GFLOPs.  This is 159 times faster than the single CPU speed,
which represents a parallel efficiency of 66\%.

\section{Results}

The initial motivation for undertaking this calculation was to
understand differences between results reported in IA and SPB.  Using
our code, which is based on the same algorithm used in SPB's
calculations, we ran a series of tests attempting to reproduce SPB's
results.  These test calculations used all of SPB's model choices,
including SPB's viscosity prescription, a Newtonian potential, and a
torus for the mass source.  We were able to reproduce most
quantitative results reported in SPB99's torus calculations.  This
includes their radial scaling laws.  For example, in a model that is
identical to SPB99's Run B, we find $\dot{M}\propto r$, $\rho\propto
r^0$, $c_s^2\propto r^{-1}$, $v_{\phi}\propto r^{-1/2}$, and
$|v_r|\propto r^{-1}$.  These power law slopes are identical to those
reported by SPB99.  As another example, we found $\dot{M} = 1.23
\times 10^{-3}$ torus masses per torus orbit at the pressure maximum
for a model identical to SPB's Model A (their fiducial model); SPB
report $\dot{M} = 1.0 \times 10^{-3}$ in the same units.  Given the
fluctuations in mass accretion rate, our value and SPB's value are
fully consistent.  We were even able to reproduce certain numerical
artifacts associated with the inner radial boundary, such as a density
drop and temperature spike near the inner boundary.

Recall that SPB evolve an initial torus and allow it to accrete; IA
use a different experimental design in which matter is steadily
injected onto the grid.  They also use a different viscosity
prescription.  We ran a second series of test calculations
attempting to reproduce IA's results.  These test calculations used
all of IA's model choices, including viscosity prescription,
Newtonian potential, etc.  We were able to reproduce all of IA99's
calculations except those that include thermal conduction
(which we did not attempt to reproduce).  In each case we found that
the qualitative nature of the flow is similar to that described in
IA99.  In particular, we agree on which models are stable and
unstable and which models exhibit outflows.  We also find
qualitative agreement with their contour plots of, e.g., density
pressure, mass flux, and Mach number.  We also find qualitative
agreement with their radial run of $c_s/V_K$ and specific angular
momentum.  Our results do not agree precisely, but this is likely
due to small differences in mass injection scheme (because IA99 do
not give their $\dot{\rho}(r,\theta)$).  Finally, we can also
reproduce the radial scalings given in IA00 for their model A.

The most significant difference between the results of IA and SPB
was due to the choice of anomalous stress prescription, as might
have been anticipated.  Qualitatively, the stress components that
are included in IA and not SPB tend to smooth the flow and suppress
turbulence.  Thus SPB99's simulations result in more vigorous
convection than simulations performed by IA.  The choice of mass
supply (torus vs.  injection) also leads to a significant difference
between IA and SPB's results; this is discussed in more detail
below.

The fact that we can reproduce both SPB's and IA's results using a
single code is consistent with the hypothesis that differences
between their reported results (e.g. the lower degree of convection
reported in IA than SPB, and the differences in radial scaling laws)
is due to differences in experimental design and viscosity
prescription rather than numerical methods.  While we cannot
completely rule out the possibility that SPB, IA, and we have made
identical experimental errors, this seems unlikely.  This comparison
thus lends credibility to SPB, IA, and our numerical results.

\subsection{Fiducial Model Evolution}

We now turn from reproducing earlier viscosity models to
considering new aspects of our own models.  First, consider the
evolution of a ``fiducial'' model (Run A in Table \ref{par} and
Table \ref{result}).  The fiducial model has $r_{in}=2.7GM/c^2$,
$r_{out}=600GM/c^2$, $r_{inj}=495GM/c^2$, $\gamma=3/2$,
$\alpha=0.1$, $N_r=108$, $N_{\theta}=50$.  It uses a
pseudo-Newtonian potential, mass is supplied by injection, and the
viscosity prescription follows IA.  It was run from $t = 0$ to
$t=7.3\times 10^5 GM/c^3$.

Run A is similar to IA99's ``Model 5'', except that it uses a
pseudo-Newtonian potential.  In a statistically steady state the
flow is characterized by a quasi-periodic outflow.  Hot bubbles
form at the interface between bound (Bernoulli parameter $Be =
(1/2) v^2 + c_s^2/(\gamma-1) + \Psi < 0$) and unbound ($Be > 0$)
material.  These hot bubbles are buoyant and move away from the
black hole.  This appears to be a low-frequency, low wavenumber
convective mode (IA refer to it as a ``unipolar outflow'').  Higher
wavenumber convective modes are evidently suppressed by the
viscosity.

\begin{figure}
\plotone{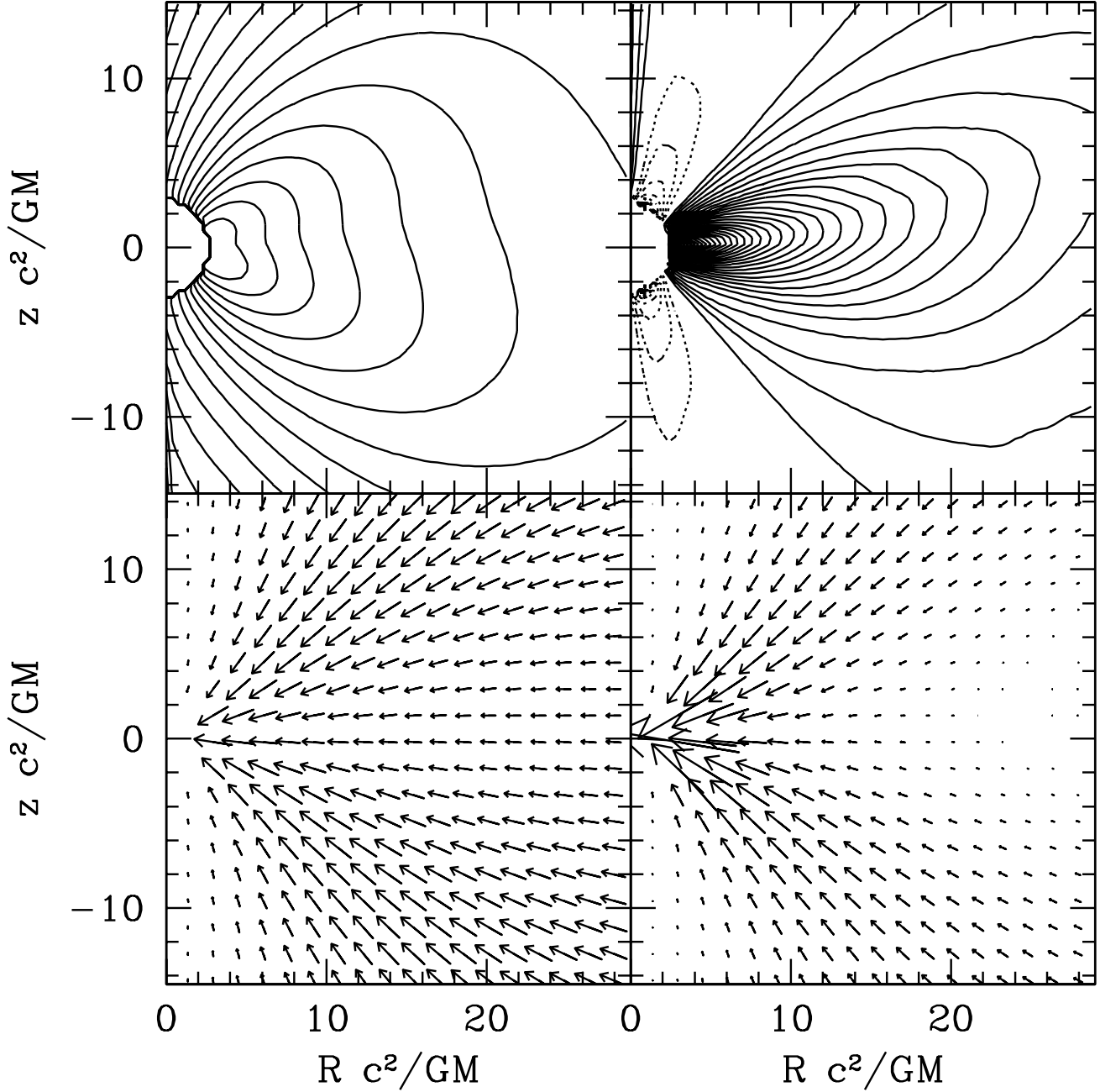}
\caption{
Time-averaged spatial structure of fiducial run (Run A; $\alpha=0.1$,
$r_{in}=2.7GM/c^2$, and $r_{out}=600GM/c^2$).  Shown are the density
(upper left), Bernoulli parameter ($Be = (1/2) v^2 +
c_s^2/(\gamma-1) + \Psi$) (upper right; dotted line is a negative
contour), scaled mass flux $r^2 \sin{\theta}(\rho\mathbf{v})$ (lower
left), and scaled angular momentum flux $r^3\sin^2{\theta}(\rho
\mathbf{v} v_{\phi}+\mathbf{\Pi}\cdot \hat{\phi})$ (lower right).  The
flow is not symmetric about the equator because the flow exhibits long
timescale antisymmetric variations.  Convective bubbles form at the
interface between positive and negative Bernoulli parameter
(i.e. unbound and bound matter).}
\label{fid-4panel}
\end{figure}

Figure~\ref{fid-4panel} shows time-averaged plots of various
quantities in the fiducial run.  The time average is performed from
200 equally spaced data dumps from $t=4.3\times 10^5 GM/c^3$ to
$t=7.3\times 10^5 GM/c^3$.  We show only the region $r_{in}<r<30
GM/c^2$, whereas the computational domain is much larger:
$r_{in}<r<600 GM/c^2$.  The injection point is located far outside
the plotted domain at $r=496 GM/c^2$.  Because of the strong
time-dependence of the flow in the fiducial run, the flow at any
instant may look very different from these time averaged plots.

The upper left panel in Figure 1 shows the average density; notice
that the density is not symmetric about the equator.  This is
because the flow involves long-timescale quasi-periodic variations
which are not quite averaged out over the course of the run.  The
upper right corner shows the Bernoulli parameter $Be$.  Dotted
lines are negative; notice that there is a substantial amount of
fluid near the equator that is unbound in the sense that $Be > 0$.
Nevertheless, this material is still flowing inward in a nearly
laminar fashion.  Near the poles, the time-averaged $Be < 0$, but
this region experiences large fluctuations.  Polar outflows are
typically associated with positive fluctuations in $Be$.  The lower
left panel shows the scaled mass flux $r^2 \sin\theta
(\rho\mathbf{v})$.  Notice that much of the mass flux is along the
surfaces of the inflow rather than along the equator.  The lower
right panel shows the scaled angular momentum flux $r^3
\sin^2\theta (\rho \mathbf{v} v_{\phi}+\mathbf{\Pi}\cdot
\hat{\phi})$.  As for the mass flux, most of the activity is along
the surface of the flow.

\begin{figure}
\plotone{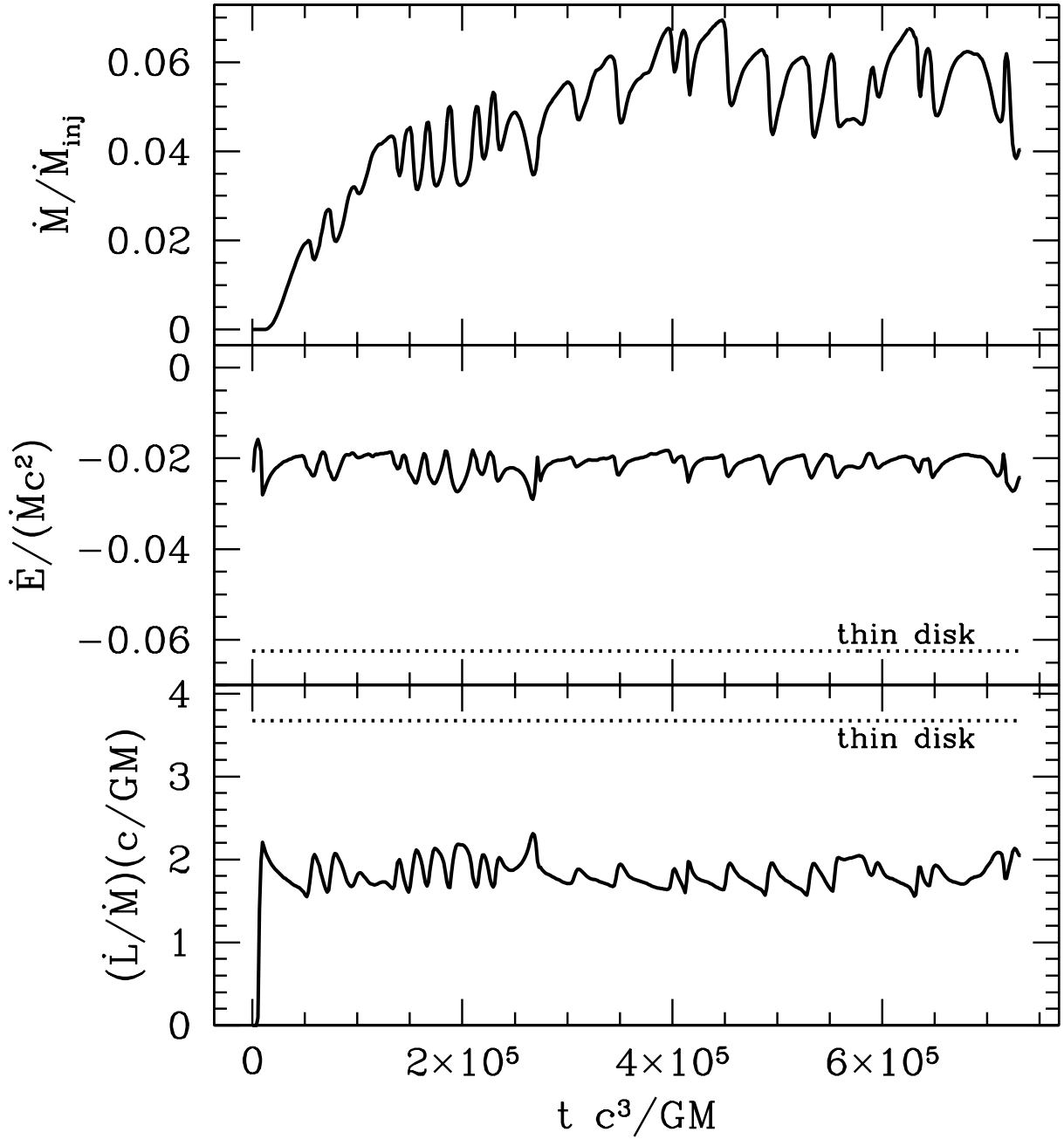}
\caption{
The evolution of $\dot{M}/\dot{M}_{inj}$, $e = \dot{E}/(\dot{M}c^2)$,
and $l = \dot{L}~c/(GM\dot{M})$ in the fiducial run (Run A).  The
dotted line indicates the thin disk value.  The run has clearly
entered a quasi-steady state.  The evolution is relatively smooth
with a small variation on a timescale $\tau\approx 4\times 10^4$.
This is the timescale for convective bubble formation (the low
point in mass accretion rate is when bubble forms).  For this model
the bubble forms at alternate poles.  A full cycle requires of
order one rotation period at the injection radius.
}
\label{fid-3dotpanel}
\end{figure}

\begin{figure}
\plotone{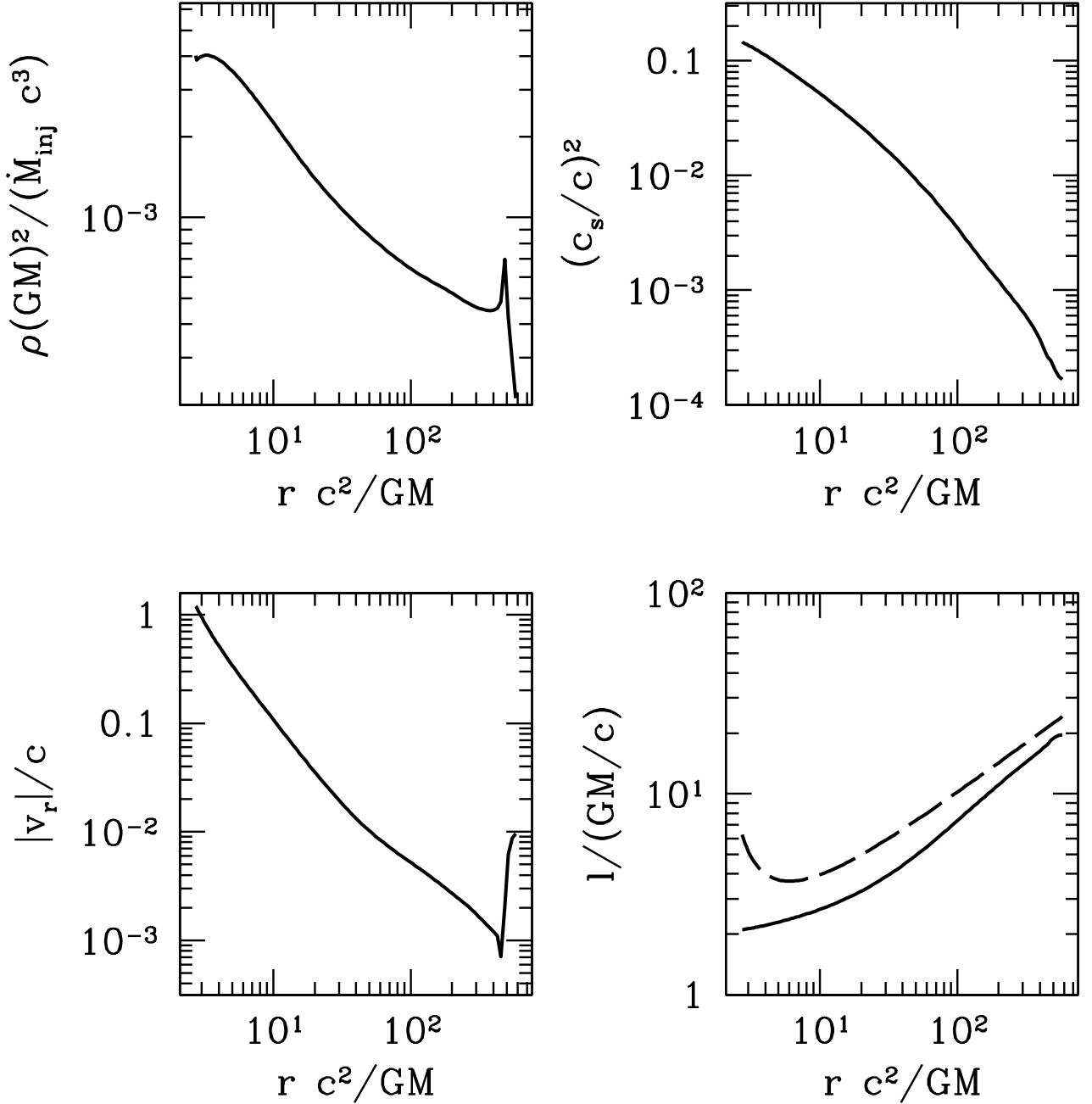}
\caption{
The radial run of $\theta$ and time averaged quantities from the
fiducial run (Run A).  Shown are the density (upper left), squared sound
speed (upper right), radial velocity (lower left), specific angular
momentum (lower right; solid line), and circular orbit specific angular
momentum (lower right; dashed line).  Crudely speaking, the inner flow
is consistent with a radial power law.  The  best fits to a power law
are: $\rho\propto r^{-0.6}$, $c_s\propto r^{-0.5}$, $|v_r|\propto
r^{-2}$, and $v_{\phi}\propto r^{-0.8}$.  The plots are averaged over
$\theta=\pi/2\pm\pi/6$.
}

\label{fid-41dpanel}
\end{figure}

Figure~\ref{fid-3dotpanel} shows the time series of the reduced
eigenvalues: $\dot{M}/\dot{M}_{inj}$, $e = \dot{E}/(\dot{M}c^2)$,
and $l = \dot{L c}/(GM\dot{M})$.  Also shown as dashed lines are the
thin disk values for $e$ and $l$.  These assume a thin, cold disk
terminating at $r = 6 GM/c^2$.  The low value of $l$ is due to two
effects.  First, the disk is already sub-Keplerian by the time the
flow reaches the innermost stable circular orbit.  In addition,
there are residual viscous torques in the plunging region that lower
the specific angular momentum of the accreted material (see
Figure~\ref{fid-41dpanel}, below).  Notice that
Figure~\ref{fid-3dotpanel} shows a smooth evolution that varies on a
timescale $\tau\approx 4\times 10^4$ at late time.  The largest
variations in mass accretion rate are related to the appearance of
large convective bubbles.

Figure~\ref{fid-41dpanel} shows the $\theta$ and time averaged run
of several quantities with radius.  The averages are taken over
$4.3\times 10^5 GM/c^3 < t < 7.3 \times 10^5 GM/c^3$ and $|\theta -
\pi/2| < \pi/6$ \footnote{Averaging over $|\theta - \pi| < \pi/36$
produces nearly identical results, but we have chosen to use IA's
range in $\theta$.}.  The upper left plot shows the run of density.
Notice that here, as for the other quantities, there is a spike
near $r_{inj}$, an intermediate region, and then an inner, roughly
power-law region.  The upper right plot shows $(c_s/c)^2$; the
lower left shows $|v_r|/c$.  Notice that the radial velocity
exceeds the speed of light at the inner boundary.  Similarly the
azimuthal velocity $v_\phi/c$ shown in the lower right panel
approaches the speed of light.  Also shown in that panel is the
circular velocity (dashed line).  Evidently the flow is slightly
sub-Keplerian at most radii.

The radial run of flow quantities in the inner regions can be fit by
power laws, as done by IA and SPB.  Our best fit power laws for the
fiducial model (Run A) over $2.7GM/c^2 < r < 20GM/c^2$ are
$\rho\propto r^{-0.6}$, $c_s\propto r^{-0.5}$, $|v_r|\propto
r^{-2}$, and $v_{\phi}\propto r^{-0.8}$.  Between $2.7GM/c^2 < r <
6GM/c^2$, $v_{\phi}$ is best fit by $v_{\phi}\propto r^{-0.9}$,
which is nearly, but not exactly, consistent with conservation of
fluid specific angular momentum ($v_{\phi}\propto r^{-1}$).  Angular
momentum is not exactly conserved at $r < 6 GM/c^2$ because of
viscous torques.

The careful reader may notice that the power law slopes quoted in
the last paragraph are not consistent with a constant mass accretion
rate.  This is because the power laws are derived from averages over
$|\theta - \pi/2| < \pi/6$, following IA.  If one averages over all
$\theta$ and time, then the resulting profiles are consistent with
constant mass, energy, and angular momentum accretion rates, as they
must be for a flow that is steady when averaged over large times.

\subsection{Dependence on Inner Boundary Location and Gravitational
Potential}

Having established that the differences between SPB and IA's models
are due to model choices rather than numerics, we were also
interested in studying whether {\it any} features of global viscous
accretion models are strongly dependent on numerical parameters.
The first parameter we considered was the location of the inner
boundary.

In models that use a Newtonian gravitational potential (such as IA
and SPB) the location of the inner boundary is not an interesting
parameter in the sense that there is no physical lengthscale that
one can compare $r_{in}$ to: it is simply a scaling parameter.  In
models that use a pseudo-Newtonian potential, however, there is a
feature (a ``pit'') in the potential on a lengthscale $G M/c^2$.
Starting with our fiducial model, then, what is the effect of
shifting $r_{in}$?

\begin{figure}
\plotone{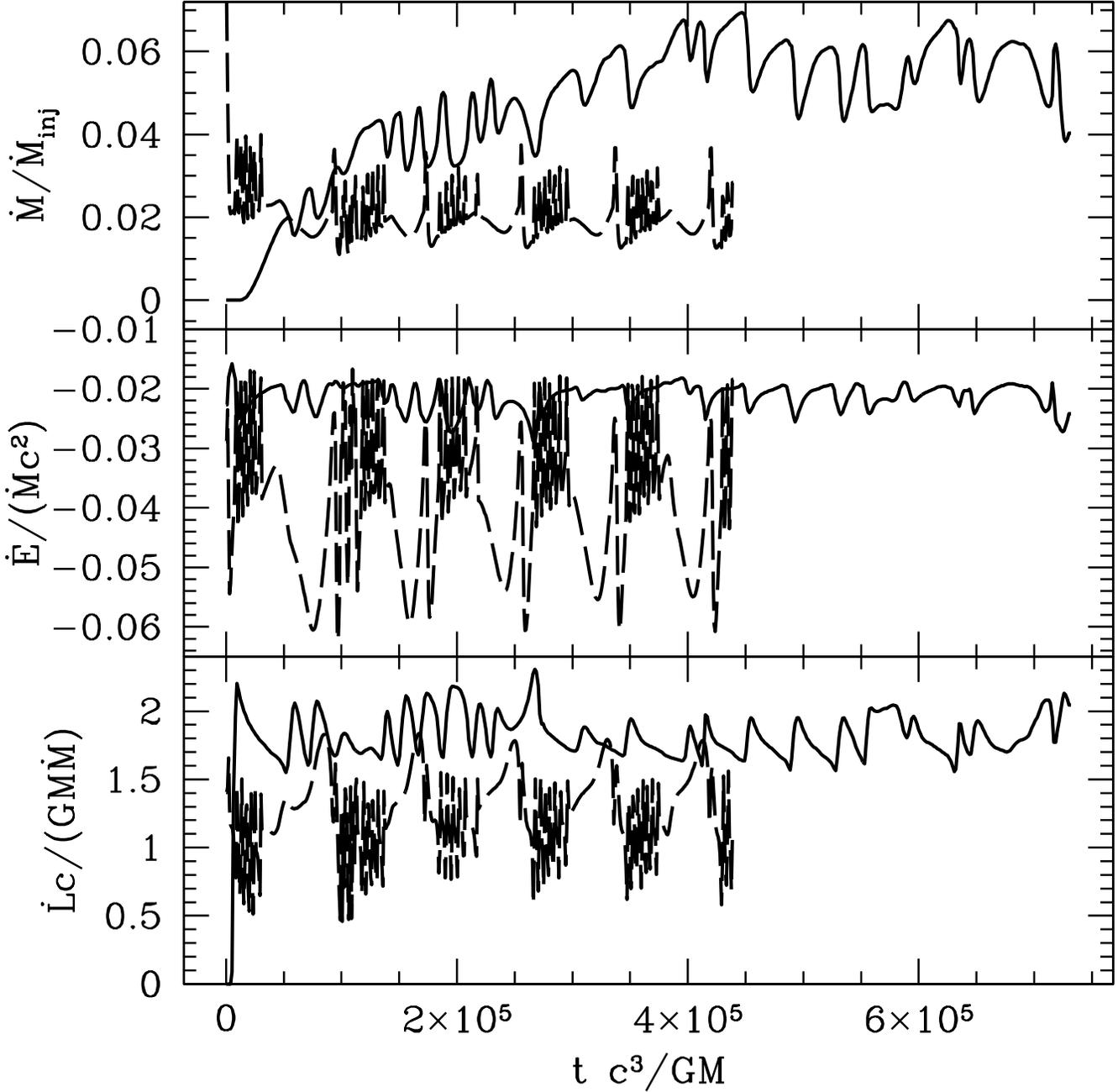}
\caption{
The effect of moving the inner boundary on the accretion rates of
mass, angular momentum, and energy (Run A vs. Run B).  The top
panel shows $\dot{M}/\dot{M}_{inj}$, the middle panel
$\dot{E}/(\dot{M}c^2)$, and the bottom panel
$\dot{L}~c/(GM\dot{M})$.  The solid curve is Run A, which has
$r_{in} = 2.7 G M/c^2$.  The dashed curve is Run B, which has
$r_{in} = 6 G M/c^2$.  Evidently Run B has a different variability
structure and different time averaged values for the accretion
rates.  The relatively rapid and high-amplitude variations in Run B
are due to nonphysical interactions with the inner radial boundary.
Only by ensuring a supersonic flow (as in Run A) can one avoid these
nonphysical effects.
}
\label{comp}
\end{figure}

Our fiducial run has $r_{in} = 2.7 G M/c^2$.  This may be
compared with Run B, which has $r_{in} = 6 G M/c^2$.  Figure
\ref{comp} compares the accretion rates in the two runs.
Evidently there are two changes in the solution.  First, the
time-averaged accretion rates differ by a large factor.  The
mean mass accretion rate is factor of $3$ lower in Run B than
Run A.  The reduced eigenvalues $e$ and $l$ also differ by about
$50$\% (see Table 2).  Second, the time variation of the
accretion rates differs, with Run B showing far more
short-timescale variations.  The short-timescale variations are
due to the interaction of unstable convective modes with the
boundary conditions.  Inspection of the runs reveals an
enhancement of convection and turbulence near $r_{in}$ in Run B.

The differences between Run B and Run A are caused by the boundary
location.  Gradual variation of $r_{in}$ (in models not discussed
in detail here) reveals that if the flow on the inner boundary is
everywhere and always supersonic, then the solution is similar to
Run A.  If the flow is subsonic, then the solution exhibits
artifacts like those seen in Run B.

Evidently forcing the flow to be supersonic on the inner boundary
causally disconnects the flow from the boundary.
\footnote{Although the viscous fluid equations of motion are not
hyperbolic, and the flow in the supersonic region is in principle
in causal contact with the rest of the flow, the coupling is
exponentially weak.}  This eliminates nonphysical reflection of linear
and nonlinear waves from the boundary and renders the precise
implementation of the numerical boundary conditions irrelevant.

We do not want the reader to think that this problem arises
because we happened to choose the wrong numerical implementation
of the boundary conditions.  Our implementation is the standard
ZEUS outflow boundary condition, and it is widely used in
astrophysical problems.  While it may be possible to implement
more transparent boundary conditions in the context of other
numerical schemes,  a survey of the numerical literature shows
that in multiple dimensions this is an area of active research
\citep{roe89, karni91, dedner01, bruneau01}, and that no general
solution to the problem has been found.

Furthermore, a simple example shows that no local extrapolation
scheme can work for all accretion problems.  Consider a
numerical model of a steady spherical inflow (Bondi flow) in a
gravitational potential $\Psi(r)$.  Let us suppose that we are
primarily interested in accurately measuring $\dot{M}$.  We know
from the analytic solution of the problem that $\dot{M}$ depends
on the shape of the potential everywhere outside the sonic
point.  If we place the inner boundary $r_{in}$ {\it outside}
the sonic point and use a local extrapolation scheme, we won't
always get the correct answer because the local extrapolation
doesn't have any information about the shape of the potential
between $r_{in}$ and the sonic point.  Put differently, one
can't determine a global solution from local extrapolation at
the boundary.  The key point is that, while aspects of the
solution may be accurate, $\dot{M}$ (and $\dot{L}$ and
$\dot{E}$) are sensitive to the boundary conditions.

It is worth noting that the outer boundary is always in causal
contact with the flow, but does not cause the same type of
artifacts as the inner boundary.  Experiments show that the flow
is {\it qualitatively} insensitive to the location and
implementation of the outer boundary condition.  The time
averaged $\dot{M}$, however, is sensitive to both $r_{out}$ and
$r_{inj}/ r_{out}$.  The time averaged $l$ and $e$ scale out
this mass dependence and so are qualitatively {\it and}
quantitatively insensitive to both $r_{out}$ and $r_{inj}/
r_{out}$.

It is also worth noting the effects of changing the
gravitational potential.  Run C (identical to IA99 Model 5) is
identical to Run B except that the potential is now Newtonian.
It is qualitatively similar to Run B, but $\dot{M}$ is now a
factor of $5$ lower than Run A.  Run C also has the property
that $l$ oscillates about $0.0$.  This is a problem if the focus
of the simulation is measuring $\dot{M}$ or $\dot{L}$.

To summarize: the location of the inner radial boundary can
determine the character of the flow.  If the flow is everywhere
and always supersonic (or super-fast-magnetosonic in MHD) on the
inner boundary then boundary-related corruption of the flow is
impossible.  Since it is computationally expensive to place the
inner boundary very deep in the potential (for our model, the
time step $dt \sim (r_{in} - 2GM/c^2)$), the optimal location
for the inner boundary is just inside the radius where the
radial Mach number always exceeds $1$.

The results of IA and SPB do not focus on the time-dependence of the
accretion values, so much of their discussion is unaffected by their
treatment of the inner radial boundary.  As discussed below, there
are small changes related to the appearance of outflows.

\subsection{Comparison of Torus and Injection Models}

The torus and injection methods represent sharply different
approaches to studying accretion flows.  The equilibrium torus
presents a physically well-posed problem, but the accretion flow
is transient: no steady state can be achieved.  The injection
method reaches a quasi-steady state, but much of the
computational domain is wasted on evolving the injection region,
which has no astrophysical analog: it is nonphysical.  It is
natural to ask whether these two widely used schemes for
supplying mass can be made comparable or used to measure any of
the same quantities.

We selected two runs, E (torus) and F (injection), that had
similar mass distributions in an evolved state.  The torus run
was studied at a time when $\dot{M}$ was close to its maximum.
Run F's $\dot{M}$ is a factor of $10$ larger than Run E's.  This
difference might have been anticipated from the sensitivity of
the injection run to $r_{out}$ and $r_{inj}/r_{out}$: runs in
which mass is concentrated closer to the outer boundary tend to
have lower accretion rates because more of the mass escapes
through the outer boundary.  The time averaged mass accretion
rate is therefore strongly dependent on the method of mass
supply.  

The energy and angular momentum accretion rates per unit mass are,
however, insensitive to the experimental design.  We find that $e$
and $l$ differ by less than 3\% in Runs E and F (see Table
\ref{par} and Table \ref{result}).  These quantities are apparently
set by conditions near the inner boundary (the ISCO for the
Pseudo-Newtonian potential), and can be measured in either type of
experiment.

\subsection{Other Parameters}

We have varied $r_{inj}$ and $r_{out}/r_{inj}$, and as
reported above, these strongly affect the time-averaged value
of $\dot{M}$.  The sense of the effect is that a simulation
with a larger $r_{out}/r_{inj}$ loses less matter through the
outer boundary, and this results in more matter streaming back
into the black hole (by up to a factor of $10$).  The
qualitative nature of the flow, however, is roughly
independent of $r_{out}/r_{inj}$ in that, e.g., the temporal
power spectrum of $\dot{M}$ is similar.  The qualitative
nature of the flow is dependent on $r_{inj}$.  If one fixes
$r_{out}/r_{inj}$ and all other parameters, the range in
$\alpha$ where unipolar outflows are observed tends to become
smaller and disappears altogether for $r_{inj}$ as small as
$40 G M/c^2$.

The dependence of accretion models similar to ours on $\alpha$
has already been investigated by IA.  They find that the flow
changes from turbulent to laminar as $\alpha$ is increased and
the higher viscosity damps modes of increasing lengthscale.
IA find that $\alpha \lesssim 0.03$ the flow is turbulent, and
for $\alpha \gtrsim 0.3$ the flow is laminar.  For $0.03 <
\alpha < 0.3$ the flow exhibits a ``unipolar'' outflow.  Our
results are in agreement with IA.  However, our models with a
pseudo-Newtonian potential and super-sonic flow at the inner radial
boundary show a slight shift in the values of $\alpha$ that
exhibit unipolar outflows.

We did find a critical value of $\alpha\approx0.5$ above which
no supersonic flow at $r_{in}$ could be achieved due to
viscous heating, at least for $\gamma=3/2$ and $\gamma=5/3$
and for $r_{in}\ge 2.1GM/c^2$.  Smaller values of $r_{in}$
were not computationally practical.  This high $\alpha$ is
typically associated with a bipolar outflow, as seen by IA.
Even in this case, however, a choice of $r_{in}=2.1 GM/c^2$
instead of $r_{in}=6 GM/c^2$ leads to a qualitatively
different profile for the flow.  The flow with smaller
$r_{in}=2.1 GM/c^2$ has a bipolar outflow starting at larger
radius ($10 GM/c^2$) rather than immediately on the boundary
as with $r_{in}=6 GM/c^2$, and the mass accretion rate
increases by a factor of $3$.

Finally, we studied the dependence of the results on numerical
resolution.  We find that $N_r \times N_\theta = 108 \times 50$
is sufficient at $\alpha = 0.1$ to resolve the shortest
wavelength convective mode.  Also, we chose our value of $r_{0}$
to agree with SPB99's torus models.  We experimented with
varying $r_{0}$ and find that, all things being equal, smaller
$r_{0}$ gives more laminar flow.  

\section{Summary}

Work in this field will shortly focus on global MHD models in
pseudo-Newtonian potentials and in full general relativity.
In our view it is useful to understand the solution space for
physically and numerically simpler viscous models before
turning to MHD.  It is even possible, as \citet{sp01} have
claimed, that viscous hydrodynamics provides a crude
approximation to the MHD results.  In any event, this
investigation provides a preview of some of the experimental
issues that will play a role in most future global numerical
investigations of accretion flows.

This investigation was initially motivated by a desire to
understand whether the differences between earlier global
viscous hydrodynamics simulations performed by IA and SPB were
caused by differences in experimental design or numerical
method.  IA and SPB reported different degrees of convective
turbulence in their models and found different radial scalings
for vertically averaged quantities such as temperature and
density.  Using a single code, we were able to reproduce both
sets of results.  We conclude that the differences are due
to experimental design.

We also found, while reproducing IA and SPB's results, that
some aspects of our solutions were sensitive to the numerical
treatment of the region close to the inner boundary in models
that use a pseudo-Newtonian potential.  In particular, $l$ and
$e$, the specific angular momentum and energy of accreted
material, are strongly dependent on $r_{in}$, the location of
the inner boundary.  When the flow is supersonic at $r_{in}$
the location of the boundary does not affect $l$ and $e$.  But
when the flow is subsonic at $r_{in}$ the flow interacts
strongly with the numerical boundary condition.  This produces
spurious outflow events and makes $l$ and $e$ dependent on
$r_{in}$.  Evidently for accurate measurement of these
quantities it is necessary to isolate the numerical boundary
condition behind a sonic transition that is located within the
computational domain; one must place the inner boundary
condition inside a ``sound horizon''.

We are not saying that all models that lack a sonic transition
in the computational domain are fatally flawed.  Whether the
treatment of the inner boundary condition is problematic or
not depends on what is being measured.  For the nonlinear
eigenvalues $\dot{L}, \dot{E}$, and $\dot{M}$ that we have
focused on here, however, the treatment of the inner boundary
condition is crucial.  Furthermore, the only guarantee that
the inner boundary condition is not governing the solution is
to isolate it behind a sonic transition; this is the only
completely safe choice.

The location of the inner boundary may prove even more crucial
in MHD models.  Much of the character of the flow is
determined in the turbulent, energetically important region of
the flow just outside the fast magnetosonic transition, just
as the region immediately outside the sonic transition
determines the nature of the viscous flows described in this
paper.  We have performed some preliminary numerical tests and
find that, as in the viscous flow, $l$ and $e$ for an MHD flow
are sensitive to the treatment of the inner boundary.  For
various reasons it may prove difficult to achieve a fast
magnetosonic transition in the computational domain; accurate
treatment of the inner boundary condition may require fully
(general) relativistic MHD.

We have also compared two commonly used experimental designs
for black hole accretion flow studies: models that begin with
an equilibrium torus, and models that continuously inject
fluid onto the grid.  The choice between these models is to
some degree a matter of taste.  We find the equilibrium torus
slightly easier to initialize and analyze.  Remarkably, the
two different approaches produce indistinguishable
measurements for $l$ and $e$, the specific angular momentum
and energy of the accreted material.

A parallel, viscous, axisymmetric hydrodynamics code based on
that used in this paper can be found at
\url{http://kerr.physics.uiuc.edu}.

\acknowledgements

This work was supported in part by a GE fellowship and NASA
GSRP Fellowship Grant NGT5-50343 to JCM, an NCSA Faculty
Fellowship for CFG, the UIUC Research Board, and NSF Grant
AST-0093091.  Computations were done in part under National
Computational Science Alliance grants AST010012N and
AST010009N using the Origin 2000 and posic Linux cluster at
NCSA.  We thank Stu Shapiro and the referee for comments
that substantially improved this paper.

\appendix
\section{Rate of Strain Tensor}

The rate of strain tensor $\mathbf{e}$ is a symmetric tensor 
that in spherical polar coordinates has
\begin{equation}
e^{rr} = \frac{\partial v_r}{\partial r} - \frac{1}{3}(\nabla\cdot v),
\end{equation}
\begin{equation}
e^{\theta\theta} = \frac{1}{r}\frac{\partial v_{\theta}}{\partial \theta} +\frac{v_r}{r} - \frac{1}{3}(\nabla\cdot v),
\end{equation}
\begin{equation}
e^{\phi\phi} = \frac{v_r}{r} + \frac{v_{\theta}}{r}\cot{\theta} - \frac{1}{3}(\nabla\cdot v)+ \frac{1}{r\sin{\theta}}\frac{\partial v_{\phi}}{\partial \phi},
\end{equation}
\begin{equation}
e^{r\theta} = \frac{1}{2}(r \frac{\partial}{\partial r}(\frac{v_{\theta}}{r}) +\frac{1}{r}\frac{\partial v_{r}}{\partial \theta}),
\end{equation}
\begin{equation}
e^{r\phi} = \frac{1}{2}(r \frac{\partial}{\partial r}(\frac{v_{\phi}}{r}) +\frac{1}{r\sin{\theta}}\frac{\partial v_{r}}{\partial \phi}),
\end{equation}
and
\begin{equation}
e^{\theta\phi} = \frac{1}{2}(\frac{\sin{\theta}}{r} \frac{\partial}{\partial \theta}(\frac{v_{\phi}}{\sin{\theta}}) +\frac{1}{r\sin{\theta}}\frac{\partial v_{\theta}}{\partial \phi}).
\end{equation}


\clearpage

\clearpage
\begin{deluxetable}{lrrrrrrrrrr}
\tabletypesize{\scriptsize}
\tablecaption{Parameter List\label{par}}
\tablewidth{0pt}  
\tablehead{
  \colhead{Run}
& \colhead{$N_r$}
& \colhead{$N_{\theta}$}
& \colhead{Visc.}
& \colhead{Potential}
& \colhead{$R_{in}/r_g$}
& \colhead{$R_{out}/r_g$}
& \colhead{$R_{inj}/r_g$}
& \colhead{$\gamma$}
& \colhead{$\alpha$}
& \colhead{$t_f(c^3/GM)$}
}
\startdata
A      & 108 &  50 & IA & PN & 1.35  & 300  & 248 & 3/2   & 0.1      & $7.3\times 10^5$  \\
B      & 80  &  50 & IA & PN & 3     & 300  & 248 & 3/2   & 0.1      & $4.4\times 10^5$  \\
C      & 80  &  50 & IA & Newt. & 3     & 300  & 248 & 3/2   & 0.1      & $7.3\times 10^5$  \\
D      & 64  &  40 & MG & PN & 1.2   &  76  & 62  & 3/2   & 1.0      & $3.3\times 10^3$  \\
E      & 128 &  80 & MG & PN & 1.4    &  21 & 17  & 5/3   & 0.01     & $3.0\times 10^4$  \\
F      & 128 &  80 & MG & PN & 1.4    &  81 & 21  & 5/3   & 0.01     & $6.8\times 10^4$  \\
\enddata
\tablecomments{
IA and MG are viscosity prescription described in equations 7-9.
PN is the pseudo-Newtonian potential of Paczynski \& Wiita
(1980).
Run B uses Run C as initial conditions.  
$r_g=2GM/c^2$.  
$R_{inj}$ for Run F is the position of the torus density peak $\rho_0$.
}
\end{deluxetable}


\clearpage
%
%
%


\begin{deluxetable}{lccccc}
\tablewidth{0pc}
\tabletypesize{\scriptsize}
\tablecaption{Results List\label{result}}

\tablehead{
\colhead{Run}
& \colhead{Steady State Time ($GM/c^3$)}
& \colhead{Max. Mach at $R_{in}$}
& \colhead{$\dot{M}$} 
& \colhead{$-(\dot{E}/(\dot{M}c^2))\times 10^{-2}$} 
& \colhead{$\dot{L}~c/(GM\dot{M})$}
}

\startdata
A     & $4.3\times 10^5$   &-1.4    & $5.96\times 10^{-2}$    & $2.06$        & $1.75$     \\
B     & $2.4\times 10^5$   &+0.0    & $1.95\times 10^{-2}$    & $3.72$        & $1.29$     \\
C     & $2.4\times 10^5$   &+0.0    & $9.46\times 10^{-3}$    & $6.77$        & $.0746$    \\
D     & $\ge3.3\times 10^3$&-0.4    & $\ge3.59\times 10^{-2}$ & $4.64$        & $-0.167$   \\
E     & $5.5\times 10^3$   &-3.0    & $3.48\times 10^{-2}$    & $3.01$        & $3.41$     \\
F     & $2.0\times 10^4$   &-3.4    & $5.03\times 10^{-1}$    & $3.11$        & $3.35$     \\ 
\enddata
\tablecomments{Runs A-E are injection runs with mass accretion rate units in $\dot{M}_{inj}$ and Run F is a torus run with mass accretion rate unit in $\rho_0(GM)^2/c^3$.  Run C's angular momentum fluctuations are $10$ times the average value shown.}
\end{deluxetable}

\end{document}